\documentclass[superscriptaddress,twocolumn,nofootinbib]{revtex4-2}

\usepackage{times,fancyhdr}
\usepackage[dvips]{graphicx}
\usepackage{color}
\usepackage{mathrsfs}
\usepackage{setspace}
\usepackage{bm}
\usepackage{dsfont}
\usepackage{amsmath,amsthm,amssymb,bm}
\usepackage{hyperref}
\usepackage{xcolor,soul}
\newcommand{\ket}[1]{\left| #1 \right\rangle}

\newcommand{\upket}{\ket{\uparrow}}
\newcommand{\downket}{\ket{\downarrow}}
\newcommand{\braket}[2]{\langle #1|#2 \rangle}

\newcommand{\ketbra}[2]{\left|#1\right\rangle\hskip-1mm\left\langle #2\right|}

\begin{document}
\title{Could wavefunctions simultaneously represent knowledge and reality?}

\author{Jonte R. Hance}
\email{jonte.hance@bristol.ac.uk}
\affiliation{Quantum Engineering Technology Laboratory, Department of Electrical and Electronic Engineering, University of Bristol, Woodland Road, Bristol, BS8 1UB, UK}
\author{John Rarity}
\affiliation{Quantum Engineering Technology Laboratory, Department of Electrical and Electronic Engineering, University of Bristol, Woodland Road, Bristol, BS8 1UB, UK}
\author{James Ladyman}
\email{james.ladyman@bristol.ac.uk}
\affiliation{Department of Philosophy, University of Bristol, Cotham House, Bristol, BS6 6JL, UK}

\begin{abstract}

In discussion of the interpretation of quantum mechanics the terms `ontic' and `epistemic' are often used in the sense of pertaining to what exists, and pertaining to cognition or knowledge respectively. The terms are also often associated with the formal definitions given by Harrigan and Spekkens for the wavefunction in quantum mechanics to be $\psi$-ontic or $\psi$-epistemic in the context of the ontological models framework. The formal definitions are contradictories, so that the wavefunction can be either $\psi$-epistemic or $\psi$-ontic but not both. However, we argue, nothing about the informal ideas of epistemic and ontic interpretations rules out wavefunctions representing both reality and knowledge. The implications of the Pusey-Barrett-Rudolph theorem and many other issues may be rethought in the light of our analysis.

\end{abstract}

\maketitle

\section{Introduction}

Quantum mechanics describes the behaviour of subatomic particles. Unlike classical mechanics, which attributes particles particular values of position and momentum, quantum mechanics attributes wavefunctions, from which can be derived probabilities for possible values of position, momentum and other observables. In the quantum case the products of the standard deviation of position and the momentum (and other pairs of incompatible observables) have a minimum value given by the uncertainty principle \cite{Kennard1927Quantenmechanik}, unlike in the classical case. However, in some ways the situation is similar to how a probability distribution in classical statistical mechanics can be interpreted as describing what is known about microstate of the particles in a gas given its macrostate, and so as `epistemic'. Quantum mechanical wavefunctions behave like classical ensembles in the limit, and so perhaps the wavefunction in quantum mechanics is epistemic in a similar way. On the other hand, Fuch's Quantum Bayesianism (QBism) treats the wavefunction as representing knowledge of possible results, rather than of the underlying microstate \cite{Fuchs2002Info,Fuchs2014QBism} and so as epistemic, but in a different way, while Ben-Menahem's interpretation holds ``quantum probabilities as objective constraints on the information made available by measurement" \mbox{\cite{Benmenahem2017PBR}}.All these responses to the peculiarities of quantum mechanics - phenomena such as contextuality, entanglement and collapse - do not posit any kind of novel metaphysics and are exclusively epistemic interpretations.

By contrast, dynamical collapse theories \cite{Ghirardi1986Unified,Penrose1996Gravity}, Bohmian mechanics \cite{Bohm1952Suggested} and Everett's relative-state interpretation \cite{Everett1957Relative}, take the wavefunction to represent something physical, and not to be epistemic in any way.\footnote{Recently an epistemic interpretation of the wavefunction in Bohmian mechanics has been proposed by \mbox{\cite{Bohumeanism}}.} An influential paper by Harrigan and Spekkens formally defines the wavefunction being $\psi$-ontic or being $\psi$-epistemic as contradictory, such that the wavefunction can only be one or the other \cite{Harrigan2010Nonlocality}. However, the informal ideas of pertaining to what exists, and pertaining to knowledge do not exclude each other, as the next section shows. Section \ref{Sect:OMF} explains Harrigan and Spekkens's formalisation, and shows how their definitions of $\psi$-ontology and $\psi$-epistemicism simply presuppose that the wavefunction cannot pertain to both knowledge and reality. If, as we argue, it is right that nothing about the informal ideas of epistemic and ontic interpretations rules out wavefunctions pertaining to both knowledge and reality, then the now standard definitions create a false dichotomy\footnote{Myrvold independently noted that taking $\psi$-epistemic to be simply the negation of $\psi$-ontic is `potentially misleading' \mbox{\cite{Myrvold2020QSRealism}} and seems to take this to be obvious. However, it is not a prominent point in his paper, and is worth arguing here since some people claim it is not just not obvious but false.}. Section \ref{Sect:Conseq} considers the consequences of this for discussions of quantum foundations.

\section{Ontic and Epistemic Interpretations of Terms in Physical Theories}
\label{sec:O&E}

The notions of epistemic and ontic have wide application in philosophy. For example, an epistemic interpretation of probability takes it to be a matter of belief, information or knowledge, while an ontic interpretation of probability takes it to be something in the world and nothing to do with the representations of agents. In this and many other cases, epistemic and ontic interpretations are taken to be mutually exclusive. In general however, `epistemic' means pertaining to cognition or knowledge, while `ontic' means pertaining to what is exists (the words are directly based on Greek genitive forms of the words for cognition or knowledge and reality respectively), and there is clearly no contradiction in predicating both of the same thing without the assumption that nothing can be both epistemic and ontic. There is no argument for such an assumption in general of which we are aware, and indeed it is arguably false in particular cases. 

For example, the thermodynamic entropy $S$ of a system is a candidate for being both ontic and epistemic. Arguably, the entropy assigned to a system depends on knowledge of it as well as how the system is objectively. For instance, the entropy assigned to an equally divided box of particles is higher if an observer can distinguish these particles than if they cannot \mbox{\cite{Cheng2009Thermodynamics}}. David Wallace \mbox{\cite{Wallace}} argues that thermodynamics should be interpreted as a `control theory' in which what agents know about systems is essential to what they can do with them. Arguably, at least, the entropy of a system represents the system as it is independently of us to some extent, while also representing what is known about it \mbox{\cite{Ladyman2008UseofITEntropy}}.

Another example is that of proper mixtures in quantum mechanics, which represent both systems and what is known (or not known) about them. For example, the (proper) mixed state $(_x\ketbra{\uparrow}{\uparrow}_x + _x\ketbra{\downarrow}{\downarrow}_x)/2$, is assigned to a system when it is really either in the state $\upket_x$ or $\downket_x$), and when the epistemic probability of state $\upket_x$ and of state $\downket_x$ is a half. Clearly, proper mixtures pertain to both what exists and what is known about it. Similarly, one can imagine another observer validly assigning the state $(2_x\ketbra{\uparrow}{\uparrow}_x + _x\ketbra{\downarrow}{\downarrow}_x)/3$, illustrating that these states are to some degree epistemic - but given they are also based on the knowledge the system is actually in one of the two eigenstates of spin-$x$ with certainty (we are just unsure which), they are also ontic.

However, the terminology of `epistemic' and `ontic' that has become standard among physicists working in quantum foundations is understood so as to make the application of the terms mutually exclusive without an argument being given that they should be. In his influential review of the PBR theorem, Matt Leifer says that an ontic state represents `something that objectively exists in the world, independently of any observer or agent' \cite{Leifer2014Review}. This is stronger than the definition above because it rules out that the state also represents our knowledge of the world. He gives the example of the ontic state of a single classical particle being its position and momentum. His corresponding example of an epistemic state is then a probability distribution over the particle's phase space. Note that in the case Leifer considers there is clearly a many-one map between the epistemic state and the underlying state, in the sense that many epistemic states are compatible with the true state of the system (because there are many probability distributions that give a non-zero probability to a given state of the particle). However, this is not so for the ontic state. Only one mathematical representation is compatible with the underlying state of the system, because the mathematical representation of each possible microstate is a complete specification of all the degrees of freedom of the system over which the probability distribution is defined. As the next section explains, Harrigan and Spekkens make this difference between a many-one and one-one relationship definitive of epistemic and ontic interpretations respectively.

Simon Friedrich says a ($\psi$-)epistemic interpretation is any view according to which ``quantum states do not represent features of physical reality, but reflect, in some way to be specified, the epistemic relations of the agents who assign them to systems they are assigned to'' \mbox{\cite{friederich2014Interpreting}}. So he (and Leifer) agree that epistemic and ontic interpretations are exclusive of each other, so that an epistemic interpretation of a term rules out that the term represents the world as well. This is so for Harrigan and Spekkens too, as explained in the next section.

    However, not everything in physics is as straightforward as the toy model Leifer considers, and it is not always clear whether terms in physical theories represent real things (for example, component versus resultant forces, gauges and so on). In quantum mechanics, individual particles do not in general have their own pure quantum states. Nothing about the idea of a determinate reality requires that all physical states are decomposable into the states of particles, and the states of quantum field theory do not take individual particles to be the fundamental bearers of physical properties. The probability distributions we get from quantum mechanics are over eigenvalues associated with observables, not over underlying states-of-the-world as in classical statistical mechanics. For all these reasons and more the framework and definitions of the next section that have become orthodox are at least questionable\footnote{As a referee of a previous version of this paper pointed out, there is a long history of interpretations of the wavefunction as both ontic and epistemic, which shows precisely why it is not appropriate to treat these two terms as mutually exclusive \mbox{\cite{Bohr1928Postulate,friederich2014Interpreting,heisenberg1958physics,schrodinger1952there}}. However, for some reason, this has been neglected by Harrigan and Spekkens, who define $\psi$-ontic and $\psi$-epistemic as contradictories.}.

Moreover, it is possible to give definitions of ontic and epistemic, which capture the informal usage of the terms reasonably well, and which do not make them exclusive of each other. Wavefunctions, like particle coordinates or vector fields, are part of the mathematical apparatus of physical theory. Such a term in a physical theory has an ontic interpretation (is an `ontic state') when it is taken to represent how the world is independently of our knowledge of it \emph{to some extent or other}.\footnote{Note that this is not to say that wavefunctions are physical things, but that they represent something physical, just as the correspondence between your fingers and the numbers one to ten does not make the numbers themselves physical \mbox{\cite{Hermens2021Howreal,Schlosshauer2012Implications,Wallace2010StateRealism}}.} For example, the number 9.81 represents the acceleration due to gravity near the surface of the Earth. It is not perfectly accurate and depends on the conventional choice of units, but arguably all representations are like this to some extent. Many mathematical terms used in classical physics, whether scalars, like mass or charge, or vectors, such as magnetic field strength or linear momentum, are apt to be taken as representing real physical properties. On the other hand, a term in physical theory has an epistemic interpretation (is an `epistemic state') when it is taken to represent knowledge or information about the world \emph{to some extent or other}. (Note that the positive second part of Friederich's definition of a ($\psi$-)epistemic interpretation is very similar to ours.)

\section{The Ontological Models Framework}\label{Sect:OMF}

\begin{figure}
    \centering
    \includegraphics[width=\linewidth]{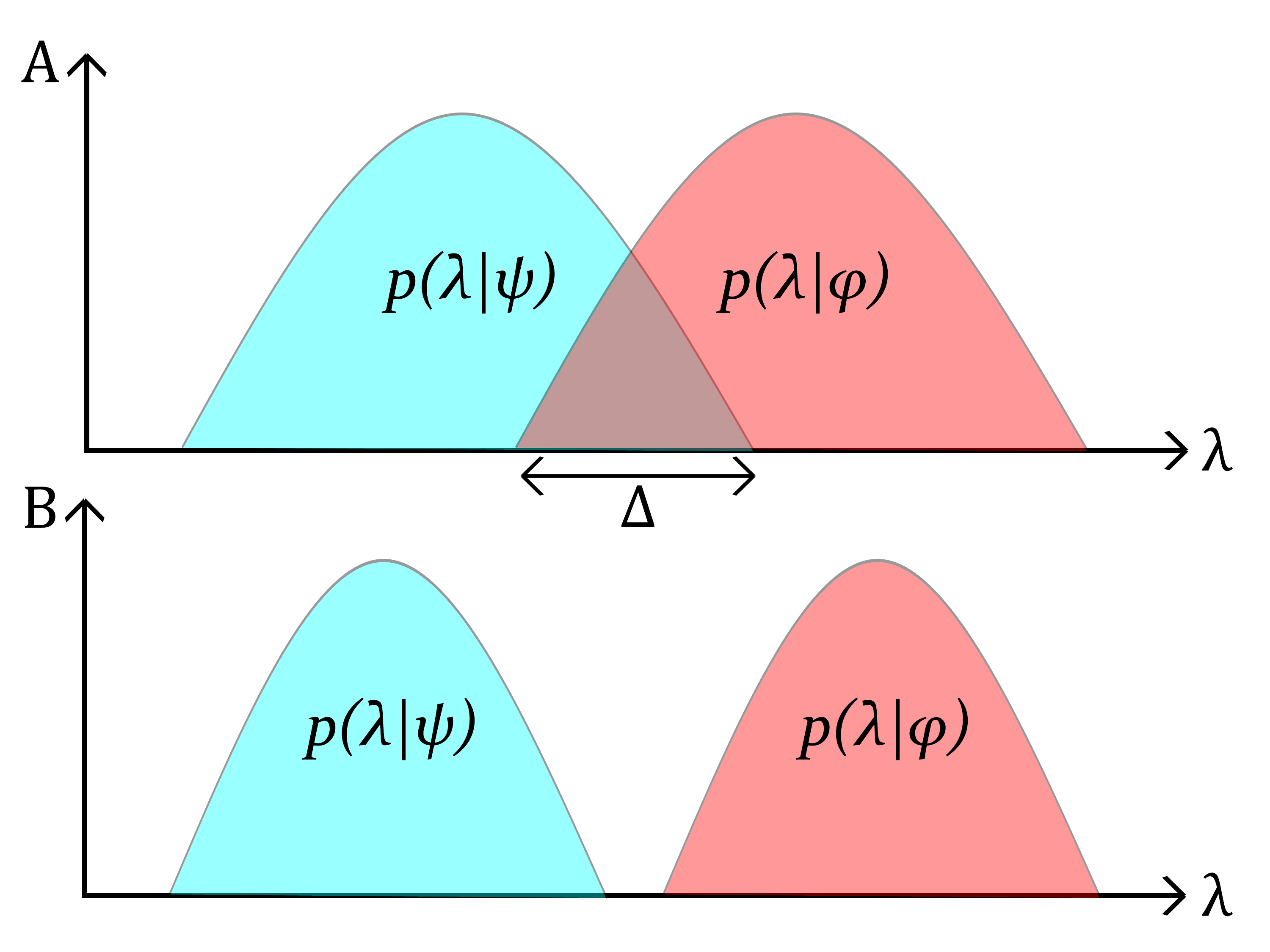}
    \caption{Harrigan and Spekkens's $\psi$-epistemic (A) and $\psi$-ontic (B) models of reality. Wavefunctions $\psi$ and $\varphi$ each have probability distributions over state-space $\Lambda=\{\lambda\}$. In a $\psi$-epistemic model, these can overlap over a subspace \textit{$\Delta$}, so a state $\lambda$ within this overlap could be represented by both $\psi$ and $\varphi$. However, in their formalism, in $\psi$-ontic models, each state can only be represented by one wavefunction.}
    \label{fig:Graphs}
\end{figure}

The wavefunction $\psi$ gives the relative (Born) probability $p^\psi_A(s)$ that, upon measurement, a certain degree of freedom (described by observable $A$) will have the specific value $s$. The wavefunction can be represented as a normalised vector $\ket{\psi}$ in a Hilbert Space $\mathcal{H}$, where there is a complete orthonormal basis of vectors, corresponding to possible values of $A$ (and we write vector $\ket{s}$ with eigenvalue $s$ for $A$). Harrigan and Spekkens's Ontological Models framework considers how the wavefunction relates to some particular underlying real state of the world $\lambda$, from the space $\Lambda$ of possible such states \cite{Spekkens2005Contextuality,Rudolph2006Ontological,Harrigan2007Ontological,Spekkens2007EpistTot, Harrigan2010Nonlocality}.

In any hidden variable model, each observable $A$ has an associated response function $\mathcal{A}(S\vert\lambda)$, which gives the probability for state $\lambda$ that a measurement of variable $A$ would give an outcome $s$ in set $S$ \cite{Schlosshauer2012Implications}. $p(\lambda\vert\psi)$ is the probability a situation described by wavefunction $\psi$ has the underlying state $\lambda$. This probability function is normalised over $\psi$'s support in state space $\Lambda_\psi$.
The framework assumes these are both probabilities (i.e., non-negative and additive) rather than quasiprobabilities (which need not be either) \cite{Harrigan2007ProbDistn}.\footnote{Note, Harrigan and Rudolph extend this framework further by including a treatment of preparation and measurement devices.} From these we get the Born rule,
\begin{equation}
\label{EqBorn}
    p^\psi_A(S)
    =\sum_S\vert\braket{s}{\psi}\vert^2
    = \int_{\Lambda}\mathcal{A}(S\vert\lambda) p(\lambda\vert\psi) d\lambda
\end{equation}

Imagining our (arbitrary) observable $A$ had the original wavefunction $\ket{\psi}$ as a possible state after measurement (i.e., $\ket{\psi}=\ket{s}$ for some $\ket{s}$ that is an eigenstate of $A$ with some value $q$), we find 
\begin{equation}
    \forall\lambda\in\Lambda_\psi,\;\mathcal{A}(q\vert\lambda) = 1 
\end{equation}
(as $\vert\braket{\psi}{\psi}\vert^2=1$, and $p(\lambda\vert\psi)$ is normalised over $\Lambda_\psi$ \cite{Maroney2012Statistical}).

If we now consider a second wavefunction, $\varphi$, with probability distribution $p(\lambda\vert\varphi)$ over its support $\Lambda_\varphi$, obeying these same rules, then, by Eq.\ref{EqBorn},
\begin{equation}
    \vert\braket{\psi}{\varphi}\vert^2
    = \int_{\Lambda}\mathcal{A}(q\vert\lambda) p(\lambda\vert\varphi) d\lambda
\end{equation}

By restricting this to the subspace $\Lambda_\psi$, we define
\begin{equation}
\begin{split}
\Delta&\equiv\int_{\Lambda_\psi}\mathcal{A}(q\vert\lambda) p(\lambda\vert\varphi) d\lambda\\
    &=\int_{\Lambda_\psi} p(\lambda\vert\varphi) d\lambda\leq\vert\braket{\psi}{\varphi}\vert^2
    \end{split}
\end{equation}

If $\Delta>0$, Harrigan and Spekkens claim this overlap of $p(\lambda\vert\varphi)$ and $\Lambda_\psi$ implies the two wavefunctions in this model are epistemic states, because the same underlying state $\lambda$ could be represented by either depending on what it known about the system.\footnote{This doesn't however mean that they necessarily represent the underlying state equally well - for a given $\lambda$ in this overlap, $p(\lambda\vert\psi)$ needn't equal $p(\lambda\vert\varphi)$.} Harrigan and Spekkens take the existence of two distinct wavefunctions with overlapping support within a model as necessary and sufficient for that model being $\psi$-epistemic (and not $\psi$-ontic). This is as, given two different wavefunctions can represent the same underlying state, the wavefunction is not determined completely by the underlying state. Therefore, some other factor must fix the wavefunction, and that factor is what is known about the system.

Harrigan and Spekkens stipulate that all models that do not meet their criterion for being $\psi$-epistemic are $\psi$-ontic, based on how, classically, states can be understood as either ontic (points of state space) or epistemic (probability distributions over state space) as discussed above.\footnote{Schlosshauer and Fine apply the terms `Mixed' and `Segregated', rather than `$\psi$-epistemic' and `$\psi$-ontic' \cite{Schlosshauer2012Implications}, which better reflects the distinction between wavefunction overlap and not.}

Harrigan and Spekkens give three subcategories of $\psi$-ontic model. In the strongest ($\psi$-complete), the wavefunction describes the system's state completely and no hidden variables are left out by it (as in, for example, Everettian interpretations). In the other two models, both referred to as $\psi$-supplemented, the state is supplemented by some hidden variable - in the first as a one-to-one mapping between the wavefunction and the state; and in the second where wavefunctions can map to more than one state (but each state still only maps to one wavefunction). While the wavefunction may not itself be the state, in $\psi$-ontic models each ontic state corresponds to at most one wavefunction (see Fig. \ref{fig:Graphs}B), so all wavefunctions have disjoint support in the state-space:
\begin{equation}
    \forall\psi,\varphi,\;s.t.\psi\neq\varphi;\;\Lambda_{\psi}\cap\Lambda_{\varphi}=\oslash
\end{equation}

\section{Analysing Harrigan and Spekkens's Definitions}\label{Sect:Analysing}

In Harrigan and Spekkens's framework, only $\psi$-epistemic models allow multiple wavefunctions to have overlapping support from the same underlying real state (see Fig. \ref{fig:Graphs}A). As noted above, they also take such overlap also to be necessary for a model to interpreted epistemically. However, while we grant that within the ontological models formalism overlap is sufficient for a model to be interpreted epistemically, we believe it is not necessary for it to be interpreted epistemically, because there is no reason to suppose that a one-one map between the wavefunction and the ontic state rules out that the wavefunction represents knowledge. Indeed, if the response functions for individual $\lambda$ within a wavefunction's support $\Lambda_\psi$ for $S$, are not equal to the Born probability for $S$ conditional on the wavefunction, then this surely suggests the wavefunction could be interpreted epistemically:

\begin{equation}
\label{Eq:EpistemicCriterion}
\begin{split}
  \exists\{\psi,S,A\},\;s.t.\;\neg(\mathcal{A}(S\vert\lambda)=p^\psi_A(S)\;\forall\lambda\in\Lambda_{\psi})\\
  \implies\;\text{epistemic}
  \end{split}
\end{equation}

On this condition the Born probability is just the weighted average of response functions over $\Lambda_\psi$, rather than equalling the response function of the state of the world $\lambda$ (suggesting there is a real difference between the $\lambda$s in the support of $\psi$, but that the experimenter does not know enough to distinguish between these different $\lambda$s)\footnote{The Colbeck-Renner-Leegwater Theorem claims to prohibit cases like this, where the quantum state does not provide a full description for the prediction of future measurement outcomes  \mbox{\cite{Colbeck2012Correspondence,Leegwater2016Impossibility}}. However, Hermens has shown this rests on a faulty assumption, and so these cases are not prohibited \mbox{\cite{Hermens2020CompletelyReal}}.}. All cases of overlap obey this condition, as within $\Lambda_\psi$, $\mathcal{A}(\psi\vert\lambda)=1$, but $p^\varphi_A(\psi)=\vert\braket{\psi}{\varphi}\vert^2$ which isn't equal to 1 if $\psi\neq\varphi$. However, cases without overlap could also obey this condition, as it applies to the response function of any observable, rather than the one represented by the operator whose eigenstates include $\psi$ whatever that is. Hence, Harrigan and Spekkens's criterion for a model being $\psi$-ontic - that wavefunctions never have overlapping support in the space of underlying states - is wrong to make it not $\psi$-epistemic by definition.

$\psi$-ontic models either take the wavefunction to represent the exact underlying state, or to represent some part of the state and so to be incomplete. What Harrigan and Spekkens ignore is the possibility that the wavefunction represents both the state of reality and the epistemic state of an observer, as with the case of a proper mixture or entropy discussed above. Hence, even though the wavefunction represents reality, wavefunctions could still have overlapping support on state space, because they could simultaneously represent different states of knowledge.

Because of this, there is nothing in Harrigan and Spekkens' formal definition for a model being $\psi$-ontic, nor a criterion we can build from their model, which is necessary or sufficient for a model to have wavefunctions representing reality. The closest we can come to is that, if the wavefunction doesn't represent knowledge, it must represent something else (and so represent reality) - but, given our condition for a model representing knowledge above is sufficient rather than necessary, there is no reason models without overlap need represent reality.

Harrigan and Spekkens admit that all models that are not $\psi$-complete could be said to have an epistemic character, especially where multiple states map onto one wavefunction, given this associates the wavefunction with a probability distribution over state space. However, they say they are instead interested only in pure quantum states - hence their definitions being based on overlap, given they claim, with pure states, ``$\psi$ has an ontic character if and only if a variation of $\psi$ implies a variation of reality and an epistemic character if and only if a variation of $\psi$ does not necessarily imply a variation of reality" \cite{Harrigan2010Nonlocality}. Hardy agrees, saying, given no overlap between the support of wavefunctions on the underlying state, we could deduce the wavefunction from $\lambda$, which he takes to mean the wavefunction would be written into the underlying reality of the world \cite{Hardy2013QStatesReal}. However, this is questionable, given that, as noted above, correlation between $\lambda$ and a wavefunction does not necessarily make that wavefunction physical, any more than the correspondence between your fingers and the numbers 1-10 makes those numbers physical \cite{Schlosshauer2012Implications} (which is why we defined wavefunction-ontic as the wavefunction representing the real state, rather than being real itself).

Further, considering the wavefunction to be ontic if and only if variation in it implies variation in the underlying state ignores that, while ontic models require that the wavefunction represent an element of reality, they need not only represent that. Even for pure states, so long as we are not considering a $\psi$-complete model, there is always some factor supplementing the wavefunction in the full underlying state of the world - so, conversely, knowledge could supplement that representation of reality to give us the wavefunction. Therefore, there is no reason for change in wavefunction to imply change in reality in an ontic model, even for pure states.

This is demonstrated by $\psi$-dependent models \cite{Schlosshauer2012Implications}. In a $\psi$-dependent model, the response functions depends on $\psi$ - individual overlap possibilities for each possible state are given by the relevant measurement probabilities, rather than being uniform. The simplest case of this is when $p_\psi$ is uniform across $\Lambda_\psi$, as, to retrieve the Born probabilities, $\mathcal{A}^\psi(S,\lambda)$ must then depend on $\psi$. As the wavefunction still to some degree reflects the underlying state, these models are ontic in our sense. However, they also allow overlap of wavefunctions on state space, making them $\psi$-epistemic. Schlosshauer and Fine give an ontic model which is also $\psi$-epistemic, also showing models can be both.

\section{Consequences}\label{Sect:Conseq}
The Pusey-Barrett-Rudolph (PBR) argument shows that wavefunctions cannot overlap on state space, but it says nothing about whether they are epistemic in the broader sense of Section~\ref{sec:O&E}. The same applies for other no-go theorems based on these definitions, such as Colbeck and Renner's \cite{Colbeck2012Correspondence}, Patra et al's \cite{Patra2013NoGo}, Hardy's \cite{Hardy2013QStatesReal}, and Ruebeck et al's \cite{Ruebeck2020Epistemic}. Pusey et al showed, given certain assumptions, that any model that is $\psi$-epistemic by Harrigan and Spekkens's criterion (i.e. allows wavefunction-overlap on state space) always contradicts quantum theory \cite{Pusey2012OnTheReality}. However, above we showed that wavefunctions being able to overlap on state space is a sufficient, rather than necessary condition for a model being epistemic. This means, even if those assumptions are valid, and so wavefunction overlap is incompatible with quantum physics, that does not rule out all epistemic models.

Even in the ontological models framework, models can be simultaneously ontic and epistemic - the wavefunction can represent both elements of reality, and knowledge about that reality. Harrigan's and Spekkens's terms, $\psi$-ontic and $\psi$-epistemic, do not formalise these informal ideas. 

In light of our analysis, it is important that people do not conflate the ideas of ontic and epistemic in general with Harrigan and Spekkens' specific definitions, as to do so confuses the debate about quantum foundations.

\begin{acknowledgements}
\textit{Acknowledgements:} We thank Hatim Salih and Karim Thebault for useful discussions. This work was supported by the University of York's EPSRC DTP grant EP/R513386/1, and the Quantum Communications Hub funded by the EPSRC grant EP/M013472/1.\\
\textit{Author Contribution Statement (CRediT):} JRH - Methodology, Writing (Original Draft, Review and Editing), Visualisation; JR - Supervision, Writing (Review and Editing), Funding acquisition; JL - Conceptualisation, Methodology, Writing (Original Draft, Review and Editing), Supervision\\
\textit{Competing Interest Statement:} The authors declare that there are no competing interests.\\
\textit{Data Availability Statement:} Data sharing not applicable to this article as no datasets were generated or analysed during the current study.
\end{acknowledgements}

\bibliographystyle{plainurl}
\bibliography{ref.bib}

\appendix

\section{Appendix - Issues with an Assumed Underlying State \texorpdfstring{$\lambda$}{lambda}}
As mentioned above, there are a number of issues with the idea of an underlying state assumed by the Ontological Models Framework. Classically, points in state space define exact values for all observables (e.g. a point in (3+3)-dimensional position-momentum state space gives the exact position and momentum of an object), so associating a real possible state-of-the-world to each point ensures all values are simultaneously single-valued. However, if we try and say a single point in a state space corresponds to a real quantum state-of-the-world, it enforces locality of hidden variables, so violates Bell's theorem. (The only major interpretation of quantum physics in which this is not the case is in Bohmian mechanics, where the state-of-the-world would pin down the quantum potential - but existence as real values would reinforce the nonlocality-vs-special-relativity issue (as information sent nonlocally would definitively be writ onto the state-of-the-world, so this nonlocal communication would be, to use Shimony's terms, action, rather than passion, at a distance \cite{Shimony1993Passion}).)

Further, there is an issue with representing a projective measurement in this form, if one assumes all values are simultaneously single-valued - given, if we assume all system evolution except measurement is deterministic (as quantum mechanics does), it implies measurement must physically disturb the system in a random way to replicate our observations. Spekkens tries to say this random disturbance could come from the initial state of the measurement device, and that if we knew this too, the whole interaction would be deterministic, but this sounds like superdeterminism-by-stealth.

Another issue, specifically with the idea of the real underlying state of the world assumed by the ontological models formalism, is that both the response function for a variable at an initial state, and the probability that a situation represented by a given wavefunction being in a given state, are real probabilities, rather than quasiprobabilities - they are always non-negative, and sum additively. This is unlike quantum `probability amplitudes', which are really quasiprobabilities - they can be negative or even imaginary, and so can interfere destructively with one another. This negative probability interference is a key part of quantum mechanics, explaining peculiar phenomena such as the Elitzur-Vaidman Bomb Detector \cite{Elitzur1993BombDet}, and counterfactual communication \cite{Salih2013CFComms,Hance2021Quantum} and imaging \cite{Hance2020CFGI}. It is unclear how the real non-negative probabilities of this real underlying state-of-the-world could replicate these effects - casting doubt on this framework.

See Oldofredi et al for additional issues with this underlying real state assumption in \cite{Oldofredi2020Classification}, and \mbox{\cite{Hance2021Ensemble}} for an alternative terminology for underlying-state models of the wavefunction.

\end{document}